\documentstyle{article}

\def\be{\begin{equation}}
\def\ee{\end{equation}}
\def\ben{\begin{eqnarray}}
\def\een{\end{eqnarray}}

\begin{document}
\centerline{\bf On Entangled Multiparticle Systems in Bohmian
Theory}

\centerline{Partha Ghose \footnote{Email: partha.ghose@gmail.com}}

%\email{parthag@vsnl.com}

\centerline{Centre for Philosophy and Foundations of Science,}
\centerline{Darshan Sadan, E-36 Panchshila Park, New Delhi
110017, India}

\begin{abstract}

Arguments are presented to show that the Bohmianian programme
cannot be implemented for entangled multiparticle systems in
general, analogous to the case of many-particle systems in
classical mechanics. We give two examples.
\end{abstract}
\section{General Argument} The three basic prescriptions of standard
Bohmian quantum theory \cite{bohm} are:

(1) take the wave function $\psi$ to be a solution of the
Schr\"{o}dinger equation,

(2) impose the guidance condition ${\bf p} = m\, d{\bf x}/dt =
{\bf \nabla}\, S$ where $S$ is the phase of the wave function
$\psi = R\, {\rm exp}(i S/\hbar)$, and

(3) choose the initial particle distribution $P_{t_0}$ such that
$P_{t_0} = \vert \psi \vert_{t_0}^2 = R^2_{t_0}$.

\noindent Given the prescriptions (1) through (3), one can prove
complete equivalence between this theory and standard quantum
mechanics by using the continuity equation for $R^2$ to show that
$P_t = R^2_t$ for all subsequent times.

While these prescriptions are self-consistent and work for single
particle and factorizable many-particle systems, it turns out that
they cannot be implemented for non-factorizable multi-particle
systems in general. The position coordinates for such systems turn
out to be non-separable, analogous to the case of the
Hamilton-Jacobi theory of many-particle systems in classical
mechanics \cite{goldstein}, and hence the Bohmian trajectories
cannot be calculated for such systems. In special cases when they
can be calculated, the trajectories turn out to be constrained,
leaving no room to choose arbitrary initial distributions to fit
quantum mechanical distributions. The problem is more severe than
in the classical Hamilton-Jacobi case because it arises already at
the level of two non-interacting particles.

 To see the point
clearly, let us consider an $N$-particle entangled system
described by the wave function \be \psi(x_1, x_2,..., x_N,t)= R
(x_1, x_2, ..., x_N,t)e^{\frac{i}{\hbar}S(x_1, x_2, ...,
x_N,t)}\label{wavfn1}\ee The guidance conditions which define the
velocities are \ben v_i = \frac{d x_i}{d t} &=&
\frac{1}{m}\frac{\partial S(x_1, x_2,..., x_N, t)}{\partial
x_i}\nonumber\\&=& f_i (x_1, x_2,..., x_N, t)\,\,\,\,\,\,\,\,i=
1,2,...,N \een The coordinates $x_i$ are not separable, and hence
the velocity equations are not separable for entangled states, for
had they been so, the wave function (\ref{wavfn1}) would have been
factorizable which by assumption it is not. Hence, the velocity
equations cannot be integrated to yield the trajectories.

Nevertheless, consider the specific example of a two-particle
non-factorizable wave function \be \psi(x_1, x_2, t) =
\frac{1}{\sqrt{N}(a+b)}[a\, e^{ip (x_1 - x_2)/\hbar} + b\, e^{-ip
(x_1 - x_2)/\hbar}]\,e^{- \frac{i Et}{\hbar}}\label{wavfn} \ee
where $\sqrt{N}$ is a normalization constant and $a$ and $b$ are
real parameters\footnote{We assume box normalization.}. This is a
solution of the Schr\"{o}dinger equation provided $E = p^2/m$.
The position probability density is given by \be R^2(x_1, x_2,
t)=\vert \psi(x_1, x_2, t)\vert^2 = \frac{1}{N (a+b)^2}[a^2 + b^2
+ 2 ab\,\, {\rm cos}\, p\, (x_1 - x_2)/\hbar]\label{probden}\ee
The important points about this probability density, as far as
this paper is concerned, are that (a) it is time independent and
stationary, (b) it depends only on $(x_1 - x_2)$ and (c) $(x_1 -
x_2)$ can take all values in the support of the wave function at
all times.

The phase of the wave function (\ref{wavfn}) is \be S(x_1, x_2,
t) = \hbar\, {\rm arctan}\, [(\frac{a - b}{a + b})\,{\rm
tan}\,p\,(x_1 - x_2)/\hbar]\,
 - E t + \eta \hbar\label{phase}\ee where $\eta = 0$ for $-\pi/2<
\theta < \pi/2$ and $\eta = \pi$ for $\pi/2< \theta < 3\pi/2$,
$\theta$ being $p\,(x_1 - x_2)/\hbar$. The Bohmian guidance
conditions are therefore \ben v_1 &=& \frac{d x_1}{d t} =
\frac{1}{m}\frac{\partial S(x_1, x_2, t)}{\partial x_1}\nonumber\\
&=& \frac{p}{m[1 + f^2(x_1 - x_2)]}[(\frac{a - b}{a + b})\,(1 +
{\rm tan}^2 p\,(x_1 - x_2)/\hbar)]\label{guid1}\\v_2 &=& \frac{d
x_2}{d t} = \frac{1}{m}\frac{\partial S(x_1, x_2, t)}{\partial
x_2}\nonumber\\ &=& -\frac{p}{m[1 + f^2(x_1 - x_2)]}[(\frac{a -
b}{a + b})\,(1 + {\rm tan}^2 p\,(x_1 - x_2)/\hbar)]
\label{guid2}\een where $f(x_1 - x_2)= (\frac{a - b}{a +
b})\,{\rm tan}\,p\,(x_1 - x_2)/\hbar$. The coordinates $x_i$ are
non-separable and so are these equations, i.e., the velocity of
each particle depends not only on its own position but also on the
position of the other. Nevertheless, in this case solutions still
exist that imply a certain constraint on the particle positions,
as will be shown now. It follows from (\ref{guid1}) and
(\ref{guid2}) that \be v_1 + v_2 = 0\label{cm} \ee which implies
\be x_1 + x_2 = \alpha\label{constraint1}\ee where $\alpha$ is an
arbitrary constant. This tells us that the centre-of-mass of the
particles is stationary. Furthermore, \be v_1 - v_2 = \frac{d
(x_1 - x_2)}{d t} = \frac{2 v}{[1 + f^2(x_1 - x_2)]}[(\frac{a -
b}{a + b})\,(1 + {\rm tan}^2 p\,(x_1 - x_2)/\hbar)\ee and hence
\ben\frac{a + b}{a - b}\,\int [{\rm cos}^2 p\,(x_1 - x_2)/\hbar+
(\frac{a - b }{a +
b})^2\, {\rm sin}^2 p\,(x_1 - x_2)/\hbar)]d(x_1 - x_2)\nonumber\\
= 2 v \int dt + \beta\label{condn}\een where $\beta$ is an
arbitrary constant. The solution is \be \frac{1}{2}(\frac{a^2 +
b^2 }{a^2 - b^2})(x_1 - x_2) + \frac{\hbar}{mv} (\frac{ab}{a^2 -
b^2})\,{\rm sin}\,2 p\,(x_1 - x_2)/\hbar= 2vt + \beta
\label{constraint2}\ee This is a constraint on the particle
positions as we will now show. Since the phase $S(x_1,x_2,t)$
(Eqn. \ref{phase}) is well defined for $x_1 - x_2 = 0$ and the
wave function (\ref{wavfn}) does not vanish at this point, this
equality must hold at some time. Let this time be $t_0$. Then,
$\beta = - 2 v t_0$, and $\alpha = 2x_{1}\vert_0= 2x_{2}\vert_0$
where $x_{i}\vert_0\,(i = 1,2)$ are the positions of the two
particles at $t = t_0$. Hence, \be \frac{1}{2}(\frac{a^2 + b^2
}{a^2 - b^2})(x_1 - x_2) + \frac{\hbar}{mv} (\frac{ab}{a^2 -
b^2})\,{\rm sin}\,2 p\,(x_1 - x_2)/\hbar= 2v(t -
t_0)\label{constraint3}\ee It is straightforward to see from this
equation that $x_1 - x_2 = 0$ is the only solution at $t = t_0$
provided $4 ab < (a^2 + b^2)$ which is satisfied if $b < a/3$.
Every pair of particles in the ensemble must satisfy this
constraint at $t = t_0$ no matter what the initial positions
$x_{i}\vert_0$ be. Hence, it is impossible to fit the probability
distribution of these trajectories to match the distribution
(\ref{probden}) at $t = t_0$ for a range of values of the
parameters $a$ and $b$ in spite of the freedom to choose the
initial positions of the particles arbitrarily.

Furthermore, the constraint (\ref{constraint3}) makes it doubtful
that a Bohmian state in phase space (which exists in Bohmian
theory \cite{holland}) is ergodic \cite{arnold}. On the other
hand, a quantum mechanical system is necessarily ergodic
\cite{frigerio,toda}.

However, the proof of equivalence between Bohmian theory and
quantum mechanics does go through for factorizable wave functions
for which constraints on the Bohmian trajectories do not exist.
Since a Bohmian state $(\psi, \{q_i\})$ is completely specified
by the wave function plus the set of hidden variables $\{q_i\}$,
it seems the only way to have a general proof of equivalence would
be to modify the guidance conditions so as to make them separable
in the coordinates.

\section{Appendix}
We will give another example of a constrained system. Consider
the two-particle wave function
\begin{equation} \psi (r_{1}, r_{2}, t) = \frac{1}{N}[ \frac{e^{i
k (r_{1A} + r_{2B})}}{r_{1A} r_{2B}} + \frac{e^{i k (r_{1B} +
r_{2A})}}{r_{1B} r_{2A}}]\,e^{-\frac{i}{\hbar} E t}\
\label{eq:wavfn}
\end{equation}
where $N$ is a normalization factor and
\begin{eqnarray}
r_{1A} &=& \sqrt{x_1^2 + (y_1 - a)^2 + z_1^2}\,\,\,\,\,\, r_{2B}
=\sqrt{x_2^2 + (y_2 + a)^2 + z_2^2}\label{coord1}\\ r_{1B} &=&
\sqrt{x_1^2 + (y_1 + a)^2 + z_1^2}\,\,\,\,\,\,r_{2A} = \sqrt{x_2^2
+ (y_2 - a)^2 + z_2^2}\label{coord2}\end{eqnarray} where the first
index $i$ (1,2) in $r_{ij}$ denotes the particle and the second
index $j$ denotes a point-like slit $A$ of co-ordinates (0,a,0) or
a point-like slit B of co-ordinates (0,-a,0) which are sources of
the two spherical waves in the $x \geq 0$ space. This wave
function is normalizable in a finite volume, analogous to the
plane wave case. This wave function is separately symmetric under
reflection in the $x$ axis ($y_i \rightarrow -y_i$) {\it and} the
interchange of the two particles $1 \leftrightarrow 2$.

The phase $S$ of the wave function (\ref{eq:wavfn}) is (in an
obvious notation)
\begin{eqnarray}
S &=& \hbar\, {\rm arctan} \frac{r_{1B}r_{2A}\, {\rm sin}k (r_{1A}
+ r_{2B}) + r_{1A}r_{2B}\, {\rm sin}k (r_{1B} +
r_{2A})}{r_{1B}r_{2A}\, {\rm cos}k (r_{1A} + r_{2B}) +
r_{1A}r_{2B}\, {\rm cos}k (r_{1B} + r_{2A})}\nonumber\\&\equiv&
\hbar\, {\rm arctan}\frac{N}{D} \label{eq:phase}
\end{eqnarray}
The Cartesian components of the Bohmian velocities of the two
particles can be computed from $S$ using
\begin{eqnarray}
v_{x_1} &=& \frac{d x_1}{d t} = \frac{1}{m}\frac{\partial
S}{\partial x_1} = \frac{1}{m}(\frac{\partial S}{\partial
r_{1A}}\frac{\partial r_{1A}}{\partial x_1} + \frac{\partial
S}{\partial r_{1B}}\frac{\partial r_{1B}}{\partial
x_1})\label{eq:vx1}\\v_{y_1} &=& \frac{d y_1}{d t}
=\frac{1}{m}\frac{\partial S}{\partial y_1} =
\frac{1}{m}(\frac{\partial S}{\partial r_{1A}}\frac{\partial
r_{1A}}{\partial y_1} + \frac{\partial S}{\partial
r_{1B}}\frac{\partial r_{1B}}{\partial
y_1})\label{eq:vy1}\\v_{z_1} &=& \frac{d z_1}{d t}
=\frac{1}{m}\frac{\partial S}{\partial z_1} =
\frac{1}{m}(\frac{\partial S}{\partial r_{1A}}\frac{\partial
r_{1A}}{\partial z_1} + \frac{\partial S}{\partial
r_{1B}}\frac{\partial r_{1B}}{\partial
z_1})\label{eq:vz1}\\v_{x_2} &=& \frac{d x_2}{d t} =
\frac{1}{m}\frac{\partial S}{\partial x_2} =
\frac{1}{m}(\frac{\partial S}{\partial r_{2A}}\frac{\partial
r_{2A}}{\partial x_2} + \frac{\partial S}{\partial
r_{2B}}\frac{\partial r_{2B}}{\partial
x_2})\label{eq:vx2}\\v_{y_2} &=& \frac{d y_2}{d t}
=\frac{1}{m}\frac{\partial S}{\partial y_2} =
\frac{1}{m}(\frac{\partial S}{\partial r_{2A}}\frac{\partial
r_{2A}}{\partial y_2} + \frac{\partial S}{\partial
r_{2B}}\frac{\partial r_{2B}}{\partial y_2})\label{eq:vy2}\\
v_{z_2} &=& \frac{d z_2}{d t} =\frac{1}{m}\frac{\partial
S}{\partial z_2} = \frac{1}{m}(\frac{\partial S}{\partial
r_{2A}}\frac{\partial r_{2A}}{\partial z_2} + \frac{\partial
S}{\partial r_{2B}}\frac{\partial r_{2B}}{\partial
z_2})\label{eq:vz2}\\
\end{eqnarray}
where
\begin{eqnarray}
\frac{\partial S}{\partial r_{1A} }&=& \hbar [1 + N^2/D^2]^{-1}
[\frac{k r_{1B}r_{2A}\, {\rm cos}k (r_{1A} + r_{2.B})+ r_{2B}\,
{\rm sin}k (r_{1B} + r_{2A}) }{D}\nonumber\\ &-& \frac{N}{D^2}(- k
r_{1B}r_{2A}\, {\rm sin}k (r_{1A} + r_{2B}) +
r_{2B}\, {\rm cos}k (r_{1B} + r_{2A}))]\label{eq:diff1}\\
\frac{\partial S}{\partial r_{2B}}&=& \hbar [1 + N^2/D^2]^{-1}
[\frac{k r_{1B}r_{2A}\, {\rm cos}k (r_{1A} + r_{2B})+ r_{1A}\,
{\rm sin}k (r_{1B} + r_{2A}) }{D}\nonumber\\ &-& \frac{N}{D^2}(-
k r_{1B}r_{2A}\, {\rm sin}k (r_{1A} + r_{2B}) + r_{1A}\, {\rm
cos}k (r_{1B} + r_{2A}))]\label{eq:diff2}
\end{eqnarray}
The expressions for $\partial S/\partial r_{1B}$ and $\partial
S/\partial r_{2R}$ are easily obtained by the replacements $A
\leftrightarrow B$ in the above expressions. These show that the
differential equations for the velocities of the two particles are
non-separable. As we have seen, this is a general feature of
many-particle entangled systems in Bohmian theory.

It is clear from the velocity equations (\ref{eq:vx1}) through
(\ref{eq:diff2}) that the equation for each particle can be
written solely in terms of its own coordinates provided
\begin{equation}
r_{1A} = r_{2B}\,\,\,\,\,\, {\rm and}\,\,\,\, r_{2B} =
r_{2A}\label{R}
\end{equation}
These are therefore `integrability conditions' for the velocity
equations, or equivalently, constraints that the trajectories
must satisfy. Trajectories that do not satisfy these constraints
do not exist.

\end{document}